\documentclass[journal]{IEEEtran}
\usepackage{amsmath,amssymb,amsfonts}
\usepackage{algorithm}
\usepackage{algpseudocode}

\usepackage{array}
\usepackage[caption=false,font=normalsize,labelfont=sf,textfont=sf]{subfig}
\usepackage{textcomp}
\usepackage{stfloats}
\usepackage{url}
\usepackage{verbatim}
\usepackage{graphicx}
\usepackage{subfig}
\usepackage{booktabs}
\usepackage{multirow} 
\usepackage{cite}
\begin{document}
	
	\title{ Adaptive Phase Shift Information Compression for IRS Systems: A Prompt Conditioned Variable Rate Framework}
	
	\author{
		Xianhua Yu, Dong Li,~\IEEEmembership{Senior Member,~IEEE}, Bowen Gu, Liuqing Yang,~\IEEEmembership{Fellow,~IEEE}, Sumei Sun,~\IEEEmembership{Fellow,~IEEE}, and George K. Karagiannidis,~\IEEEmembership{Fellow,~IEEE}
		\thanks{Part of this paper has been accepted in the 2025 IEEE Global Communications Conference (Globecom 2025) \cite{globecom}.}
		\thanks{Xianhua Yu is with the School of Electrical Engineering and Intelligentization, Dongguan University of Technology, Dongguan, China (e-mail: xianhuacn@foxmail.com).
			\par Dong Li is with the School of Computer Science and Engineering, Macau University of Science and Technology, Macau 999078, China (e-mail: dli@must.edu.mo). 
			\par Bowen Gu is with the School of Computer Science and Technology, Xinjiang University, Urumqi, Xinjiang 830049, China, and with Xinjiang Multimodal Intelligent Processing and Information Security Engineering Technology Research Center, Urumqi, Xinjiang 830049, China (e-mails:  bwgu@xju.edu.cn).
			\par Liuqing Yang is with the Internet of Things Thrust and Intelligent Transportation Thrust, The Hong Kong University of Science and Technology (Guangzhou), Guangzhou 510000, China, and also with the Department of Electronic and Computer Engineering, The Hong Kong University of Science and Technology, Hong Kong SAR, China (e-mail: lqyang@ust.hk). 
            \par Sumei Sun is with the Institute of Infocomm Research, Agency for Science, Technology and Research, Singapore 138632 (e-mail: sunsm@i2r.a-star.edu.sg).
			\par George K. Karagiannidis is with the Department of Electrical and Computer Engineering, Aristotle University of Thessaloniki, Greece. (e-mail: geokarag@auth.gr).}
		
	}
				%
                %
	
	
	\maketitle
	\pagestyle{empty}
	\thispagestyle{empty}
	\begin{abstract}
		Intelligent reflecting surfaces (IRSs) have become a vital technology for improving the spectrum and energy efficiency of forthcoming wireless networks. Nevertheless, practical implementation is obstructed by the excessive overhead associated with the frequent transmission of phase shift information (PSI) over bandwidth-constrained control lines. Current deep learning-based compression methods mitigate this problem but are constrained by elevated decoder complexity, inadequate flexibility to dynamic channels, and static compression ratios. This research presents a prompt-conditioned PSI compression system that integrates prompt learning inspired by large models into the PSI compression process to address these difficulties. A hybrid prompt technique that integrates soft prompt concatenation with feature-wise linear modulation (FiLM) facilitates adaptive encoding across diverse signal-to-noise ratios (SNRs), fading kinds, and compression ratios. Furthermore, a variable rate technique incorporates the compression ratio into the prompt embeddings through latent masking, enabling a singular model to adeptly balance reconstruction accuracy. Additionally, a lightweight depthwise convolutional gating (DWCG) decoder facilitates precise feature reconstruction with minimal complexity.   Comprehensive simulations indicate that the proposed framework significantly reduces NMSE compared to traditional autoencoder baselines, while ensuring robustness across various channel circumstances and accommodating variable compression ratios within a single model. These findings underscore the framework's promise as a scalable and efficient solution for real-time IRS control in next-generation wireless networks.
	\end{abstract}
	

	\section{Introduction}
	The move towards beyond fifth-generation (B5G) and sixth-generation (6G) networks is propelled by nascent applications including extended reality (XR), autonomous driving, and the Internet of Things (IoT), which necessitate ultra-low latency, high reliability, and pervasive connectivity~\cite{BGu,CWang,XShen,XCheng}.  Intelligent reflecting surfaces (IRS) have garnered considerable interest as cost-effective and energy-efficient solutions that reconfigure wireless propagation to satisfy stringent requirements without the need for additional active radio frequency (RF) chains. 
	
	An IRS often comprises a substantial array of passive or low-power components that may dynamically modify the phase of incoming electromagnetic waves, facilitating the controlling reflection of wireless signals.  By adeptly adjusting these phase shifts, Intelligent Reflecting Surfaces (IRSs) can augment link quality, broaden coverage, and increase both spectral and energy efficiency, all without elevating transmission power or hardware complexity~\cite{DLi1,DLi2,HXie1,HXie2,DLi3,DLi4,HXie3,XLi,CZhou,KLi,TWu,JHe}.  This adjustable propagation feature establishes IRSs as a crucial facilitator for sustainable and intelligent wireless networks.
	
	To fully harness the potential of IRS-aided systems, precise control of the phase shifts given to each IRS constituent is required.  This control relies on the dependable and effective transmission of phase shift information (PSI), which can be expressed as a matrix detailing the quantized phase values, represented as bit sequences for each reflecting element, transferred from the base station (BS) to the IRS controller.  The PSI delivery procedure is limited by the restricted bandwidth of the control link between the base station and the IRS controller~\cite{XYu1,XYu2}.  
	In IRS arrays with hundreds or thousands of elements, broadcasting the entire PSI results in substantial signaling overhead, using a considerable amount of available system resources.  
	The issue intensifies in dynamic contexts, such as mobile or time-varying channels, where regular PSI updates are necessary to maintain maximum performance.  These problems collectively highlight the imperative for efficient, low-overhead, and adaptable PSI compression methodologies to facilitate practical and scalable IRS implementations.

	\subsection{Related Works and Motivations}
	
	To mitigate the PSI overhead barrier, current research has investigated deep learning-based compression techniques, including autoencoder architectures that generate compact latent representations by encoding the PSI at the base station and reconstructing it at the IRS controller.  The S-GAPSCN framework~\cite{XYu2} employs an asymmetric encoder-decoder architecture, whereas PSCDN~\cite{XYu3} incorporates a denoising module within its decoder to enhance resilience to channel noise.  The adaptive compression framework ACFNet~\cite{ZLi} and the knowledge-assisted autoencoder PSFNet~\cite{HFeng} enhance reconstruction accuracy via adaptive learning and knowledge-driven feature inference.
	
	Notwithstanding these advancements, some essential issues persist.  
	(i) \textbf{Computational Complexity:}  In contrast to the BS, which possesses ample processing capabilities, the IRS controller functions under stringent hardware limitations, requiring lightweight decoders with minimal latency and a compact memory footprint to facilitate real-time reconstruction.  
	(ii) \emph{Restricted adaptability:}  Traditional deep models are trained under particular channel circumstances and suffer significant performance decline in unfamiliar situations.  Adapting to new contexts generally necessitates retraining with extensive, scenario-specific datasets, resulting in significant storage and computational burdens.  
	(iii) \emph{Constant compression ratio:}  Most current techniques are designed for a singular compression ratio, necessitating the storage of several models or retraining for varying ratios.  The model switching introduces extra latency and may render the PSI obsolete, so undermining IRS control precision.
	
	Conversely, deep learning has shown considerable promise in tackling essential difficulties at the wireless physical layer, including channel estimation, CSI feedback, signal detection, and beam management~\cite{JGuo1,HHuang,XYu4}.  The recent rise of big AI models (LAMs), especially large language models (LLMs), has prompted their application to wireless communication jobs.  For instance, CSI-GPT amalgamates generative Transformers with federated tuning for Channel State Information acquisition, whereas LLM4CP utilizes pre-trained Large Language Models for channel prediction via lightweight fine-tuning.  Likewise, numerous research have investigated beam prediction by tokenizing wireless signals into sequential representations that are modeled autoregressively.  These papers collectively demonstrate how LAMs enhance deep learning for robust and generalizable modeling in dynamic communication contexts.
	
	Nonetheless, the majority of current methodologies depend on direct fine-tuning or static embedding modification and fail to utilize the adaptability of prompt-based conditioning.  To our knowledge, a recent study~\cite{JGuo2} on CSI feedback examined prompting by appending fixed embeddings to the model input; although a valuable initial effort, this approach is deficient in dynamic adaptation and contextual awareness.  Furthermore, traditional soft prompt-based approaches, while parameter-efficient for large models~\cite{Lester2021,Li2021,Jia2022}, predominantly operate at the semantic embedding level, frequently utilizing largely static backbones, and consequently lack an explicit mechanism to control the statistical characteristics of hidden features.  In wireless systems affected by stochastic fading (e.g., Rayleigh/Rician), operational factors such as SNR, fading parameters (e.g., $K$-factor), and channel type cause significant alterations in internal feature statistics, resulting in covariate shift between training and deployment conditions.  Consequently, relying solely on semantic prompting is challenging in addressing these alterations, potentially leading to prompt feature mismatches and fragile out-of-distribution generalization.

	
	Moreover, previous studies have predominantly concentrated on CSI acquisition, while the issue of PSI overhead in IRS-assisted systems remains significantly under-researched.  The PSI contrasts with the CSI in that the IRS controller possesses limited computational capability, rendering it incapable of calculating the PSI based on the feedback from the CSI, making direct PSI preferable.  Moreover, current autoencoder-based methodologies are generally learned under static channel conditions and a constant compression ratio, which significantly limits generalization in time-varying contexts.
	
	These observations inspire this endeavor.  We intend to create a generic conditioning framework that integrates semantic and statistical adaptation for PSI compression.  This study is, to our knowledge, the first investigation into the PSI overhead dilemma in IRS-aided systems utilizing a big model-inspired, prompt-conditioned framework.  We propose a hybrid prompt-conditioned mechanism that enhances soft prompting through a feature-wise linear modulation (FiLM) statistical pathway, facilitating hierarchical semantic statistical adaptation for effective PSI compression.  This cohesive semantic statistical framework integrates huge model adaptability with effective signal compression, facilitating scalable and real-time IRS management.
	\par Note that, compared to the conference version of this work in \cite{globecom}, we newly propose a FiLM + soft-prompt hybrid prompting design with encoder-only placement, analyze convergence and complexity in greater depth, and provide more diverse and comprehensive simulation results.

	\subsection{Contributions}
	
	The main contributions of this work are summarized as follows:
	\begin{itemize}
		\item \textbf{First hybrid prompt-based PSI compression framework:}
		This is the first study to address the PSI overhead problem in IRS-aided systems through a large model inspired, prompt conditioned framework that unifies semantic prompting and statistical modulation.
		
		\item \textbf{Hybrid prompt mechanism for adaptive PSI compression:} We propose a novel hybrid prompting mechanism that explicitly couples two complementary conditioning pathways:  
		(i) a FiLM-based modulation branch, which injects context dependent scale and shift parameters into intermediate layers to normalize activation statistics with respect to instantaneous SNR, fading type, and target compression ratio, and (ii) a soft prompt token branch, which concatenates learnable prompt tokens with feature embeddings to provide semantic guidance and global attention biasing. This semantic statistical synergy enables the encoder to jointly align global structure and local feature distributions, achieving robust adaptation to dynamic wireless environments.
		
		\item \textbf{Variable-rate compression with contextual control:}
		The target compression ratio is introduced as a conditioning variable, enabling the encoder to dynamically adjust latent dimensions and quantization precision. This allows a single model to operate over multiple compression ratios, thereby achieving flexible accuracy efficiency trade-offs without retraining or model switching.
		
		\item \textbf{Lightweight decoder with DWCG design:}
		To meet the stringent hardware constraints of IRS controllers, we design a lightweight decoder based on the proposed depthwise convolutional gating (DWCG) module. By dynamically modulating local spatial features, DWCG selectively emphasizes informative elements while suppressing redundant ones, achieving high reconstruction fidelity with significantly reduced parameters and inference latency.
		
		\item \textbf{Comprehensive evaluation and analysis:}
		Extensive simulations under both static and dynamic channel conditions validate the framework’s robustness and flexibility. The results demonstrate superior performance compared with benchmark schemes. Ablation studies verify the benefits of hybrid prompting and DWCG, while complexity analysis confirms that these gains are achieved with minimal computational overhead.
	\end{itemize}

	\subsection{Organization}
	
	The next parts of this paper are structured as follows.  Section II presents the system model for IRS-assisted wireless communications and formulates the associated PSI overhead minimization problem.  Section III delineates the suggested prompt-conditioned PSI compression framework, encompassing the comprehensive architecture, the hybrid prompt mechanism featuring soft tokens and FiLM modulation, the streamlined DWCG decoder, and the variable rate compression method.  Section IV presents extensive simulation results and performance analysis that confirm the resilience, flexibility, and scalability of the proposed system across various channel circumstances.  Ultimately, Section V ends the work and delineates prospective avenues for further research.

	\subsection{Notations}
	
	Scalars, vectors, and matrices are denoted by italic, bold lowercase, and bold uppercase letters (e.g., $x$, $\mathbf{x}$, $\mathbf{X}$), respectively. The transpose and Hermitian transpose are denoted by $\mathbf{X}^T$ and $\mathbf{X}^H$, respectively. The operator $\mathrm{diag}(\mathbf{x})$ forms a diagonal matrix from vector $\mathbf{x}$; $\|\cdot\|$ denotes the Euclidean norm; and $\odot$ represents the Hadamard (element-wise) product. The complex Gaussian distribution with mean $\mathbf{m}$ and covariance $\mathbf{C}$ is denoted by $\mathcal{CN}(\mathbf{m}, \mathbf{C})$, while the real Gaussian distribution with mean $\mu$ and variance $\sigma^2$ is denoted by $\mathcal{N}(\mu, \sigma^2)$. The expectation operator is denoted by $\mathbb{E}[\cdot]$.

	\section{System Model and Problem Formulation}
	
	\subsection{System Model}
	\begin{figure}[t]
		\centering
		\includegraphics[width=0.9\linewidth]{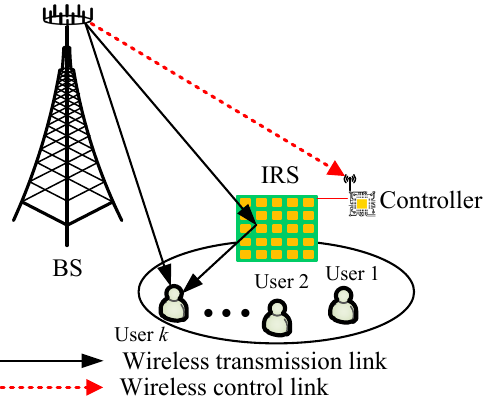}
		\caption{A typical downlink IRS-aided MIMO system.}
		\label{IRS}
	\end{figure}
	
	We consider a downlink IRS-aided multiple input multiple output (MIMO) system that consists of a single BS, an IRS, and \(K\) users, as illustrated in Fig.~\ref{IRS}. The BS is equipped with \(N_t\) antennas, and each user terminal has \(N_r\) antennas. The IRS consists of \(M\) passive reflecting elements, each of which can independently adjust the phase shift of the incident electromagnetic signal.
	
	Let \(\mathbf{G} \in \mathbb{C}^{M \times N_t}\), \(\mathbf{H}_{r,k} \in \mathbb{C}^{N_r \times M}\), and \(\mathbf{H}_{d,k} \in \mathbb{C}^{N_r \times N_t}\) denote the BS–to–IRS, IRS–to–user-\(k\), and direct BS–to–user-\(k\) channels, respectively. The BS transmits a signal vector \(\mathbf{s} \in \mathbb{C}^{N_t \times 1}\) with total power \(P\), assuming \(\mathbb{E}[\mathbf{s}\mathbf{s}^H] = \mathbf{I}_{N_t}\). The received signal at user \(k\) is given by
	\begin{equation}
		\mathbf{y}_k = \sqrt{P}\big(\mathbf{H}_{r,k}\mathbf{\Phi}\mathbf{G} + \mathbf{H}_{d,k}\big)\mathbf{s} + \mathbf{u}_k,
	\end{equation}
	where \(\mathbf{u}_k \sim \mathcal{CN}(\mathbf{0}, \sigma_k^2 \mathbf{I}_{N_r})\) denotes additive white Gaussian noise (AWGN). The IRS reflection matrix is expressed as
	\begin{equation}
		\mathbf{\Phi} = \rho\,\mathrm{diag}\!\left(e^{j\theta_1}, e^{j\theta_2}, \dots, e^{j\theta_M}\right),
	\end{equation}
	where \(\rho \in (0,1]\) denotes the reflection–amplitude coefficient (typically close to unity), and \(\theta_m \in [0, 2\pi)\) is the phase shift applied by the \(m\)-th IRS element.
	
	Owing to limited phase-resolution hardware, each \(\theta_m\) is uniformly quantized into \(B\) bits, i.e.,
	\begin{equation}
		\mathcal{Q} = \Big\{0, \tfrac{2\pi}{2^B}, \dots, (2^B-1)\tfrac{2\pi}{2^B}\Big\}.
	\end{equation}
	The quantized phase shift vector is denoted as \(\mathbf{t} = [\theta_1, \dots, \theta_M]^T \in \mathcal{Q}^M\).
	
	\textit{Remark:} The above formulation describes the data transmission link between the BS, IRS, and users. However, the PSI vector \(\mathbf{t}\) must also be delivered from the BS to the IRS controller through a separate control link with limited bandwidth, which is subject to compression, transmission noise, and distortion. The design of efficient PSI compression and reconstruction for this control link constitutes the focus of this work.
	
	\subsection{Problem Formulation}
	
	Our objective is to reliably deliver the quantized PSI vector $\mathbf{t} \in \mathcal{Q}^M$ from the BS to the IRS controller over a bandwidth-limited control link, while minimizing reconstruction error subject to channel noise and compression constraints.  
	
	Given an encoder function $f_{\text{enc}}(\cdot)$ and a decoder function $f_{\text{dec}}(\cdot)$, the encoded representation of $\mathbf{t}$ is transmitted through a noisy control channel characterized by a linear channel matrix $\mathbf{H}_c$ and additive Gaussian noise $\mathbf{w} \sim \mathcal{N}(\mathbf{0}, \sigma_w^2 \mathbf{I})$. The reconstructed PSI vector at the IRS controller is therefore given by
	\begin{equation}
		\hat{\mathbf{t}} = f_{\text{dec}}\!\left( \mathbf{H}_c f_{\text{enc}}(\mathbf{t}) + \mathbf{w} \right).
	\end{equation}
	
	The design problem can thus be formulated as minimizing the average normalized mean square error (NMSE) between the original and reconstructed PSI vectors:
	\begin{equation}
		\begin{aligned}
			\min_{\Theta} \quad & 
			\mathbb{E}\!\left[ \frac{\|\mathbf{t} - \hat{\mathbf{t}}\|^2}{\|\mathbf{t}\|^2} \right] \\
			\text{s.t.} \quad & 
			\hat{\mathbf{t}} = f_{\text{dec}}\!\left( \mathbf{H}_c f_{\text{enc}}(\mathbf{t}) + \mathbf{w} \right), \\
			& \mathbf{t} \in \mathcal{Q}^M, \\
		\end{aligned}
	\end{equation}
	where $\Theta$ denotes the learnable parameters of the encoder decoder pair, $d$ represents the latent dimension, and $r \in (0,1]$ defines the target compression ratio, i.e., the maximum allowable signaling overhead. 
	
	\subsection{Main Challenges}
	
	Despite its potential to reduce overhead in IRS-aided systems, PSI compression still faces several unresolved challenges that hinder its real time and scalable deployment:
	
	\begin{itemize}
		\item \textbf{Computational constraints at the IRS controller:}  
		The IRS controller is typically equipped with limited computational and memory resources, making it unable to support complex neural architectures. An asymmetric design with a lightweight decoder is therefore required to balance reconstruction accuracy and computational efficiency.
		
		\item \textbf{Adaptability to dynamic wireless environments:}  
		Wireless channels exhibit substantial variations across SNR levels, fading types, and spatial configurations. Conventional autoencoder based compression schemes, trained under static conditions, often experience severe performance degradation in mismatched environments because retraining with large scale, scenario specific datasets is impractical.
		
		\item \textbf{Flexibility in compression ratio:}  
		Existing schemes generally employ a fixed compression ratio, which limits their deployment flexibility. In practice, varying channel conditions and system requirements necessitate adjustable compression rates to balance reconstruction accuracy, latency, and signaling overhead without retraining or model switching.
	\end{itemize}
	
	These challenges collectively underscore the need for an efficient and adaptive PSI compression framework that can accommodate channel dynamics, support variable rate operation, and remain practical for deployment on resource-constrained IRS controllers.
	
	\section{Proposed Method}
	This section reviews autoencoder-based PSI compression and justifies the design of the asymmetric encoder-decoder architecture.  We subsequently present the prompt embedding modules and elucidate their role in conditioning the encoder for adaptive compression.  We present a variable rate mechanism utilizing prefix masking combined with energy normalization, facilitating seamless performance trade-offs without the need for retraining.  We present the lightweight DWCG decoder and the unified end-to-end training procedure that jointly optimizes encoder adaptability and decoder efficiency.
	
	We propose a novel prompt-conditioned, encoder-only framework for PSI compression in IRS-aided wireless systems.  The design addresses two practical requirements:  Adaptability to dynamic wireless environments and support for variable compression ratios within a single unified model.  A prompt is generated at the BS from side information \(s=\{\mathrm{SNR},\,\mathrm{channel\ type},\,r\}\), where \(r\) represents the target compression ratio, and is utilized to condition the encoder during feature extraction.  The prompt is introduced solely into the encoder via two complementary mechanisms: (i) soft prompt tokens that offer global semantic guidance and (ii) FiLM, which facilitates local feature calibration.  The decoder is designed to be prompt-free and lightweight in order to meet the computational and bandwidth limitations of the IRS controller.
	
	\subsection{Preliminaries}
	
	We first review the three building blocks that underpin our proposed method: autoencoder based PSI compression, asymmetric encoder decoder design, and attention-enhanced feature extraction.
	
	1) Autoencoder based PSI compression.
	The PSI is represented as a rasterized map \(\mathbf{T}\in\mathbb{R}^{H\times W}\) corresponding to the spatial layout of IRS elements, with \(M=HW\) total entries. Its vectorized form is denoted by \(\mathbf{t}=\mathrm{vec}(\mathbf{T})\in\mathbb{R}^{M}\). The encoder compresses \(\mathbf{t}\) into a compact latent representation \(\mathbf{z}\in\mathbb{R}^{d}\), and the decoder reconstructs it as
	\begin{equation}
		\mathbf{z}=f_{\mathrm{enc}}(\mathbf{t}), \qquad
		\hat{\mathbf{t}}=f_{\mathrm{dec}}(\mathbf{z}).
	\end{equation}
	
	2) Asymmetric encoder decoder design.
	Since the IRS controller is highly resource constrained, we adopt an asymmetric architecture comprising a moderately complex BS side encoder and a lightweight IRS side decoder, consistent with practical deployment scenarios. The latent representation is transmitted over a bandwidth-limited and noisy control link modeled as
	\begin{equation}
		\tilde{\mathbf{z}}=\mathbf{H}_c\mathbf{z}+\mathbf{w},
	\end{equation}
	where \(\mathbf{H}_c\) represents linear impairments such as gain mismatch, quantization error, or diagonal fading, and \(\mathbf{w}\sim\mathcal{N}(\mathbf{0},\sigma_c^2\mathbf{I})\) denotes additive noise. The IRS controller reconstructs \(\hat{\mathbf{t}}=f_{\mathrm{dec}}(\tilde{\mathbf{z}})\). This asymmetric setup motivates the prompt at encoder and prompt free decoder paradigm introduced in the following sections.
	
	3) Attention-enhanced feature extraction.
	To effectively capture long range correlations among IRS elements, the encoder incorporates self-attention layers as part of its backbone. Given input token features \(\mathbf{X}\in\mathbb{R}^{L\times d}\), multi-head attention (MHA) computes
	\begin{equation}
		\mathrm{MHA}(\mathbf{X})
		= \big[\mathrm{Attn}_1(\mathbf{X});\,\dots;\,\mathrm{Attn}_H(\mathbf{X})\big]\mathbf{W}_O,
	\end{equation}
	where each attention head is defined as
	\begin{equation}
		\mathrm{Attn}_h(\mathbf{X}) = \mathrm{softmax}\!\left(\frac{\mathbf{Q}_h\mathbf{K}_h^\top}{\sqrt{d_h}}\right)\mathbf{V}_h,
	\end{equation}
	with \(\mathbf{Q}_h=\mathbf{X}\mathbf{W}_{Q,h}\), \(\mathbf{K}_h=\mathbf{X}\mathbf{W}_{K,h}\), and \(\mathbf{V}_h=\mathbf{X}\mathbf{W}_{V,h}\).
	Residual connections, layer normalization, and position wise feed forward layers are further employed to stabilize training and improve generalization. These attention blocks form the encoder backbone that will later be conditioned by prompts for adaptive compression under varying channel and rate conditions.

	\subsection{Prompt Conditioning Mechanism}
	\begin{figure*}[t]
		\centering
		\includegraphics[width=1\linewidth]{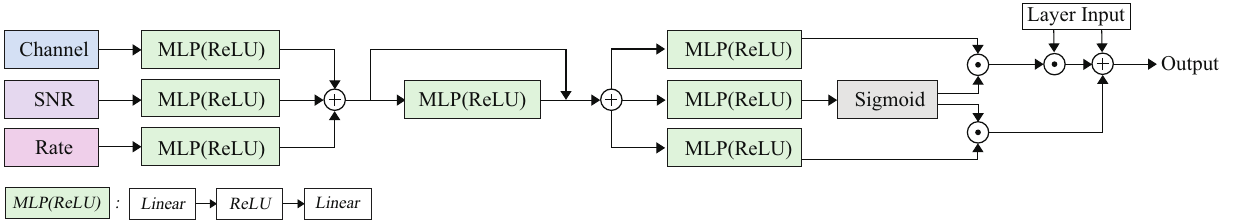}
		\caption{FiLM-based local feature modulation.}
		\label{film}
	\end{figure*}
	
	\begin{figure*}[t]
		\centering
		\includegraphics[width=1\linewidth]{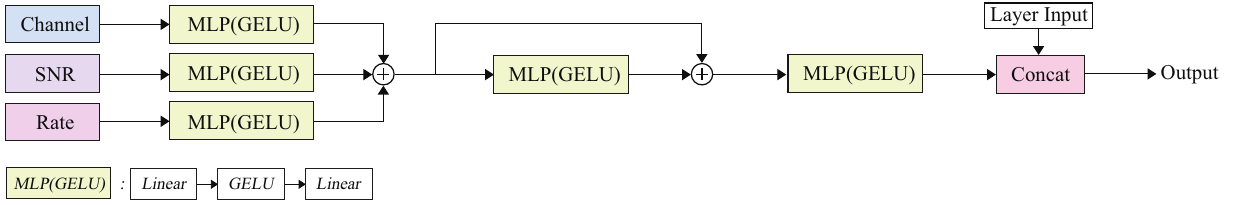}
		\caption{Soft prompt semantic token injection.}
		\label{soft}
	\end{figure*}
	Current PSI compression networks are generally trained under static channel conditions and a singular compression ratio, resulting in significant sensitivity to dynamic wireless environments.  Static encoder–decoder pairs frequently experience significant reconstruction degradation or inefficient bit usage when there are changes in fading statistics, SNR, or control-link bandwidth.  Despite the proposal of various adaptive schemes, including hyper network-based and retraining-based designs, these approaches are impractical for real-time intelligent reflecting surface (IRS) control. This is primarily due to the constraints of resource-limited hardware, which cannot accommodate frequent updates to the decoder.
	
	Recent designs of soft prompts primarily function at the semantic embedding level, utilizing learnable tokens to direct attention; however, they do not incorporate a clear mechanism for adjusting the statistical properties of intermediate features.  As a result, they face challenges in accommodating physical variations (e.g., SNR, fading parameters, channel type) and are susceptible to distribution mismatch when channel statistics change.  Furthermore, prompts generated at the base station from the channel state information are transmitted to the IRS-side decoder, which involves traversing the wireless channel and may result in prompt distortion and latent misalignment.  These limitations necessitate a local encoder that employs a statistically informed prompting approach.
	
	This motivates an encoder-only, prompt-conditioned, and context-aware adaptation mechanism that achieves real-time flexibility without modifying the decoder.  Instead of modifying network weights or sending explicit side information to the decoder, the base station creates a lightweight prompt based on the current channel context and target compression ratio, \(s=\{\mathrm{SNR},\,\mathrm{ChanType},\,r\}\), and incorporates it exclusively into the encoder, leaving the decoder free of prompts to prevent over-the-air prompt distortion and prompt–latent mismatch.  As demonstrated in Fig.~\ref{film} and Fig.~\ref{soft}, the prompt directs the encoder’s feature extraction through two complementary approaches: (1) \textbf{FiLM-based feature modulation}, which executes \emph{fine-grained statistical calibration} by introducing context-dependent scale and shift coefficients into intermediate layers, and (2) \textbf{soft prompt token injection}, which offers \emph{semantic guidance} through token concatenation prior to the attention layers.  These pathways facilitate parameter-efficient, context-aware adaptation in response to dynamic channel conditions and varying compression ratios.
	
	\noindent\textbf{Overview of the Principle.}
	The prompt conditioning mechanism functions as an adaptive control interface that connects side information to the encoder feature space.  Figures \ref{film} and \ref{soft} demonstrate that the side information \(s=\{\mathrm{SNR},\mathrm{ChanType},r\}\) (representing \textit{Channel}, \textit{SNR}, and \textit{Rate} inputs) is converted into a low-dimensional prompt representation, which captures instantaneous channel statistics and target rate constraints.  This prompt adjusts the encoder in two synchronized phases.  In the FiLM branch (Fig.~\ref{film}), the prompt adjusts intermediate activations by scaling and shifting them to align with the statistical profile of the current channel (e.g., mean/variance), thus ensuring the consistency of latent statistics.  In the soft prompt branch (Fig.~\ref{soft}), semantic tokens are integrated through concatenation (\textit{Concat}) with the layer input to direct self-attention towards spatial spectral structures that are most pertinent to the instantaneous SNR and compression ratio.  This dual FiLM soft prompt modulation enables the encoder to be entirely self-adaptive during inference, generating representations that align with current link conditions without altering decoder weights.
	
	\noindent\textbf{1) FiLM-Based Feature Modulation.}
	FiLM performs fine-grained statistical calibration of intermediate features using side information.
	A ReLU MLP maps $s=[\mathrm{SNR},\,r,\,\mathbf{c}_{\text{chan}}]$ to a descriptor $\mathbf{p}_f$, which is projected to per-layer, per-channel scale/shift:
	\begin{equation}
		\begin{aligned}
			\boldsymbol{\gamma}_\ell&=\mathrm{Proj}_{\gamma,\ell}(\mathbf{p}_f),\quad \\
			\boldsymbol{\beta}_\ell&=\mathrm{Proj}_{\beta,\ell}(\mathbf{p}_f),\quad\\
			\mathbf{H}'_\ell&=\mathbf{H}_\ell\odot(1{+}g_f\boldsymbol{\gamma}_\ell)+g_f\boldsymbol{\beta}_\ell,
		\end{aligned}
	\end{equation}
	with $g_f=\sigma(\mathbf{w}_f^\top\mathbf{p}_f)$ controlling the modulation strength (channel wise broadcasting implied). $\boldsymbol{\beta}_\ell$ recenters and $\boldsymbol{\gamma}_\ell$ rescales activations to match target statistics.
	Under low SNR/high compression, informative channels can be amplified while noise dominated channels are attenuated, at high link quality, dynamics are compressed to avoid saturation.
	Ignoring cross channel correlations and noting that shifts do not change variance,
	\[
	\mathrm{Var}[H'_{\ell,c}]\;\approx\;(1{+}g_f\gamma_{\ell,c})^2\,\mathrm{Var}[H_{\ell,c}],
	\]
	showing that FiLM learns when to amplify or damp specific channels rather than enforcing a fixed increase/decrease.
	This stabilizes latent distributions and reduces statistical drift across operating regimes.
	
	\medskip
	\noindent\textbf{2) Soft Prompt Token Injection.}
	Soft prompts act at the semantic attention level.
	A GELU MLP maps the same $s$ to a semantic descriptor $\mathbf{p}_s$, which is projected into $P$ prompt tokens:
	\begin{equation}
		\mathbf{S}=\mathrm{Proj}_s(\mathbf{p}_s)\in\mathbb{R}^{P\times d},\qquad
		\tilde{\mathbf{X}}_{\text{in}}=[\mathbf{S};\,\mathbf{X}_{\text{in}}].
	\end{equation}
	These tokens enter self attention with the data tokens so that global correlations are guided by channel/rate priors.
	In shorthand, with $Q_X,K_X,V_X$ from data and $Q_S,K_S,V_S$ from prompts,
	\[
	\mathrm{Attn}(\tilde{\mathbf{X}}_{\text{in}})=\mathrm{softmax}\!\Big(
	\frac{\begin{bmatrix}Q_S\\ Q_X\end{bmatrix}
		\begin{bmatrix}K_S^\top & K_X^\top\end{bmatrix}}{\sqrt{d}}\Big)
	\begin{bmatrix}V_S\\ V_X\end{bmatrix}.
	\]
	The cross terms $Q_XK_S^\top$ behave like a \emph{low rank prior bias} on attention scores, steering focus toward structures consistent with $(\mathrm{SNR},\mathrm{ChanType},r)$, while $Q_SK_X^\top$ lets prompt tokens act as anchors that pool context-matching evidence.
	During training, gradients update both prompt embeddings and attention weights, boosting informative structure and suppressing redundancy.
	In terms of cost, soft prompts only extend the sequence length in the first attention layer from $N$ to $N{+}P$, adding about $Pd$ parameters (we use $P\!\le\!8$).
	
	\begin{figure*}[t]
		\centering
		\includegraphics[width=1\linewidth]{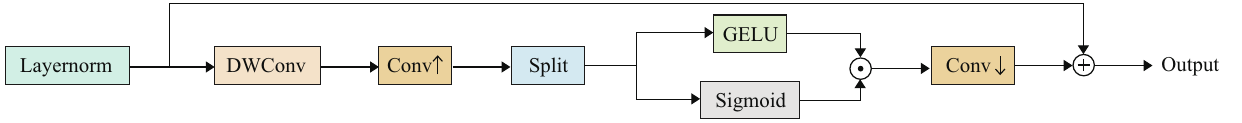}
		\caption{Structure of the proposed DWCG block.}
		\label{DWCG}
	\end{figure*}
	
	\noindent\textbf{3) Hierarchical and Rate-Aware Synergy.}
	Our design implements a hierarchical conditioning pipeline: FiLM first normalizes feature statistics with respect to the physical channel, and soft prompts then reshape global attention according to contextual semantics, enabling robust operation across diverse SNRs, channel types, and compression ratios without retraining or decoder changes. The two conditioning paths regulate complementary aspects of the encoder representation. FiLM modulates statistical characteristics to stabilize feature distributions, whereas soft prompts adjust the attention geometry to emphasize structure relevant features. This complementary action reduces coupling between pathways and stabilizes convergence. The compression ratio \(r\) is jointly embedded in both branches, allowing the encoder to anticipate effective bandwidth and align its latent representation with instantaneous channel capacity. This hierarchical semantic statistical conditioning enables real-time, encoder-only adaptation without retraining or decoder modification.
	
	\noindent\textbf{4) Complexity and Overhead.}
	Both conditioning branches introduce limited computational overhead. For an encoder with $L$ layers and feature width $d$, FiLM adds approximately $2Ld$ per-layer scale shift parameters plus a small gating module, resulting in $\mathcal{O}(Ld)$ additional FLOPs. soft prompt injection adds \(P \times d\) learnable parameters and increases only the first self-attention’s sequence length from \(N\) to \(N{+}P\), contributing an extra \(\mathcal{O}(PNd)\) FLOPs (ignoring small \(P^2\) terms; approximately linear in \(P\) for fixed \(N\)). Overall, the prompt conditioning enables a single model to sustain high reconstruction fidelity across diverse SNRs, fading conditions, and compression ratios.

	\subsection{DWCG for Lightweight PSI Reconstruction}
	
	A central design principle of the proposed framework is encoder decoder asymmetry. Prompt conditioning is applied exclusively at the encoder, while the decoder remains prompt free. This design aligns with the deployment realities of IRS-aided systems: the BS can support high complexity adaptive processing, whereas the IRS controller operates under stringent computational and memory constraints.  
	Transmitting prompt embeddings through the narrow control link would undermine the intended compression gains; therefore, the decoder focuses solely on efficient reconstruction of the prompt-encoded latent representation.
	
	To this end, we propose the DWCG architecture, which achieves high fidelity PSI recovery with linear complexity. DWCG replaces self attention, whose computational cost grows quadratically with sequence length $\mathcal{O}((HW)^2C)$, with depthwise convolutions that capture localized dependencies at $\mathcal{O}(HWC)$ complexity. These are further enhanced with dynamic gating that adaptively reweights spatial features, enabling selective feature emphasis while suppressing redundancy. This design offers a favorable balance between efficiency and expressiveness, supporting near real time inference on low computational resource IRS controllers without compromising reconstruction accuracy.
	
	\noindent\textbf{1) DWCG Block Architecture.}
	As illustrated in Fig.~\ref{DWCG}, each DWCG block processes an input sequence $\mathbf{X}\!\in\!\mathbb{R}^{B\times T\times E}$ in three stages: depthwise convolution, gated activation, and residual fusion.  
	An optional layer normalization is first applied to stabilize activations:
	\begin{equation}
		\tilde{\mathbf{X}} = \mathrm{LayerNorm}(\mathbf{X}).
	\end{equation}
	Local dependencies are then captured through a depthwise convolution with kernel size $k$:
	\begin{equation}
		\mathbf{H} = \mathrm{DWConv}_k(\tilde{\mathbf{X}}).
	\end{equation}
	A subsequent pointwise convolution expands the channel dimension by an expansion factor $\eta$ and divides the output into a value branch and a gate branch:
	\begin{equation}
		[\mathbf{V},\,\mathbf{G}] = \mathrm{Conv1D}_{2\eta E\times E}(\mathbf{H}).
	\end{equation}
	The value branch undergoes GELU activation, while the gate branch passes through a sigmoid function to regulate feature flow:
	\begin{equation}
		\mathbf{Y} = \mathrm{Conv1D}_{E\times\eta E}\!\big(\mathrm{GELU}(\mathbf{V}) \odot \sigma(\mathbf{G})\big).
	\end{equation}
	Finally, the gated features are fused with the residual input to preserve gradient flow:
	\begin{equation}
		\mathbf{X}_{\text{out}} = \mathbf{X} + \mathbf{Y}.
	\end{equation}
	
	This operation can be viewed as a convolutional analogue of self attention: the gating branch functions as a local key value interaction that dynamically modulates spatial information flow, enhancing salient PSI patterns while suppressing irrelevant responses. Unlike attention mechanisms that require global correlation computation, DWCG models fine grained spatial coherence at strictly linear complexity, making it ideally suited for low computational resource IRS controllers.
	
	\noindent\textbf{2) Properties and Design Insights.}
	The proposed DWCG module achieves an effective balance between expressiveness and efficiency. Depthwise kernels capture localized spatial correlations among neighboring IRS elements, while the gating mechanism dynamically reweights their contributions based on contextual cues embedded in the latent representation.  
	This adaptive modulation enables the network to focus computational resources on salient regions of the PSI map, enhancing reconstruction accuracy without additional complexity. In contrast to attention based decoders designed primarily for global semantic modeling DWCG functions as a local refinement module that reconstructs fine grained phase textures while preserving the high level structural coherence established by the prompt conditioned encoder. This complementary design ensures efficient, high fidelity PSI reconstruction under the asymmetric encoder decoder paradigm.
	\begin{figure*}[t]
		\centering
		\includegraphics[width=1\linewidth]{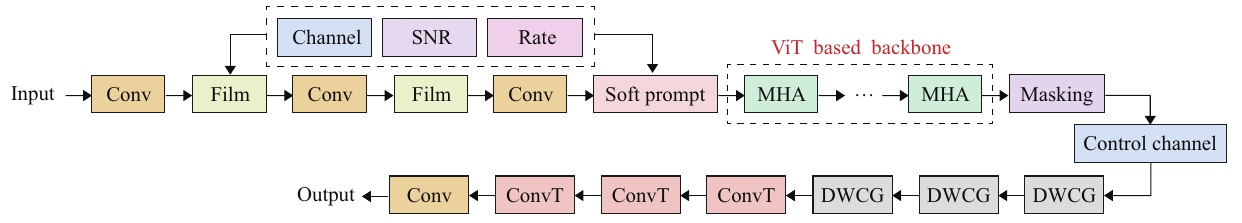}
		\caption{Overall architecture of the proposed unified encoder decoder framework.}
		\label{framework}
	\end{figure*}
	
	\noindent\textbf{3) Complexity and Complementarity.}
	For an input of size $H\times W\times C$, the computational cost of one DWCG block is $\mathcal{O}(kHWC + \eta HWC)$, scaling linearly with spatial resolution.  
	In our implementation, replacing transformer style attention with DWCG reduces decoder FLOPs by more than 80\% and the parameter count by approximately 65\%, enabling real time PSI reconstruction on embedded IRS hardware with negligible accuracy loss. Functionally, DWCG complements the encoder’s prompt guided global adaptation by providing lightweight local refinement, forming a cohesive asymmetric autoencoder that jointly optimizes adaptability, efficiency, and deployability for practical IRS-aided PSI delivery.
	
	\subsection{Unified Encoder--Decoder Framework for Variable-Rate PSI Delivery}
	
	We formulate a unified, asymmetric autoencoder with a prompt-conditioned encoder (FiLM + soft prompt) and a prompt-free DWCG decoder for adaptive PSI compression and reconstruction. 
	As illustrated in Fig.~\ref{framework}, the system comprises three tightly coupled modules:  
	(i) a dual path, prompt-conditioned encoder that embeds channel and rate awareness into the latent representation;  
	(ii) a prefix based latent masking unit enabling variable rate operation via selective activation of latent dimensions, and  
	(iii) a lightweight DWCG decoder that performs efficient, high fidelity PSI reconstruction on resource limited IRS controllers.  
	
	This joint design sustains consistent performance across heterogeneous channel types, SNR regimes, and compression ratios without retraining or architectural changes. By harmonizing encoder side adaptability with decoder side efficiency, the framework offers a practical and scalable solution for real-time PSI delivery in IRS-aided wireless networks.
	
	\noindent\textbf{1) Encoder Processing.}
	Given a PSI map $\mathbf{T}\!\in\!\mathbb{R}^{H\times W}$, the encoder first extracts low-level spatial features via a convolutional embedding layer,
	\begin{equation}
		\mathbf{X}_{\text{in}} = \mathrm{EmbedConv}(\mathbf{T}),
	\end{equation}
	yielding tokenized features $\mathbf{X}_{\text{in}}\!\in\!\mathbb{R}^{N\times d}$.  
	Side information $s=\{\mathrm{SNR},\,\mathrm{ChanType},\,r\}$ is then injected through two hierarchical conditioning branches:
	
	\begin{itemize}
		\item \textbf{FiLM-based local calibration:}  
		The FiLM module generates per-layer modulation parameters $(\boldsymbol{\gamma}_\ell,\,\boldsymbol{\beta}_\ell,\,g_f)$ that adapt activation statistics to instantaneous channel conditions.  
		The first two convolutional layers are modulated as
		\begin{equation}
			\mathbf{H}_\ell \leftarrow \mathbf{H}_\ell \odot (1+g_f\boldsymbol{\gamma}_\ell) + g_f\boldsymbol{\beta}_\ell, 
			\quad \ell\!\in\!\{1,2\}.
		\end{equation}
		This fine grained calibration maintains stable feature variance across SNR and fading regimes. We modulate only the first two convolutional layers to stabilize early stage statistics and minimize encoder latency overhead.

		\item \textbf{Soft prompt global adaptation:}  
		After FiLM calibration, a soft prompt network produces semantic tokens $\mathbf{S}\!\in\!\mathbb{R}^{P\times d}$, which are concatenated with data tokens:
		\begin{equation}
			\tilde{\mathbf{X}}_{\text{in}} = [\mathbf{S};\,\mathbf{X}_{\text{in}}].
		\end{equation}
		These prompt tokens participate in MHA, steering global feature correlations according to the embedded side information and the target compression ratio.
	\end{itemize}
	
	\noindent\textbf{ViT-based latent encoding.}  
	The conditioned sequence $\tilde{\mathbf{X}}_{\text{in}}$ is processed by a Vision Transformer (ViT)~\cite{ViT} backbone with $L$ stacked MHA blocks and feed-forward sublayers.  
	This backbone captures long range spatial dependencies among IRS elements while integrating the semantic bias introduced by soft prompts.  
	The final latent vector is obtained as
	\begin{equation}
		\mathbf{z} = f_{\mathrm{enc}}(\mathbf{T};s) \in \mathbb{R}^{D},
	\end{equation}
	where $\mathbf{z}$ encapsulates both the PSI’s spatial structure and channel rate semantics, providing a unified basis for variable-rate compression and adaptive reconstruction.
	
	\noindent\textbf{2) Variable-Rate Control via Prefix Masking.}
	To support flexible compression ratios within a single model, the latent is adaptively truncated according to the target rate $r$ (see “Masking” and “Control channel” in Fig.~\ref{framework}):
	\begin{equation}
		\mathbf{z}_r = \alpha(r)\big(\mathbf{M}(r)\odot \mathbf{z}\big),
	\end{equation}
	where $\mathbf{M}(r)\in\{0,1\}^{D}$ retains the first $k=\max(\lfloor rD\rfloor,1)$ coefficients, and $\alpha(r)=\sqrt{D/k}$ preserves latent energy to avoid rate-induced scale drift. This prefix based masking ensures that lower rate settings reuse the most informative components of higher rate ones, maintaining semantic and structural consistency across rates.  
	Because $r$ is jointly embedded in both FiLM and soft prompt branches, the encoder learns to hierarchically rank latent channels by semantic importance during training, yielding monotonic and predictable rate distortion behavior without retraining or model switching. The prompt is not signaled, only the masked latent $\mathbf{z}_r$ is transmitted through the noisy control link:
	\begin{equation}
		\tilde{\mathbf{z}} = \mathbf{H}_c\,\mathbf{z}_r + \mathbf{w},
	\end{equation}
	where $\mathbf{H}_c$ denotes the control link fading matrix and $\mathbf{w}$ is additive Gaussian noise.
	
	\noindent\textbf{3) Prompt-free DWCG Decoder.}
	At the IRS controller, the received latent $\tilde{\mathbf{z}}$ is decoded by a lightweight DWCG-based network,
	\begin{equation}
		\hat{\mathbf{T}} = f_{\mathrm{dec}}^{\mathrm{DWCG}}(\tilde{\mathbf{z}}).
	\end{equation}
	The decoder expands $\tilde{\mathbf{z}}$ via a fully connected projection, followed by a sequence of DWCG blocks that progressively restore spatial correlations and local phase continuity. By omitting prompt injection and attention operations, the decoder reduces computational cost substantially, supporting real-time inference on embedded IRS controllers while preserving reconstruction accuracy. Functionally, the DWCG decoder complements the encoder’s global, prompt-guided adaptation by performing lightweight local refinement with linear complexity $\mathcal{O}(HWC)$, reconstructing fine-grained PSI details while maintaining the global coherence established by the encoder.
	
	\noindent\textbf{4) End-to-End Training Objective.}
	The encoder and decoder are trained jointly under randomized channel, SNR, and rate conditions to minimize the normalized mean-squared error (NMSE):
	\begin{equation}
		\mathcal{L}(\Theta) =
		\mathbb{E}_{\mathbf{T},s}\!\left[\frac{\|\mathbf{T}-\hat{\mathbf{T}}\|_2^2}{\|\mathbf{T}\|_2^2}\right],
	\end{equation}
	where $\Theta$ denotes all trainable parameters of the encoder–decoder pair.  
	Two complementary sampling strategies are adopted:  
	(i) fixed rate training for per-rate benchmarking and ablation, and (ii) uniform rate training for mixed-rate generalization.  
	Stochastic rate sampling compels the encoder to hierarchically organize latent dimensions by semantic importance, forming a rate-ordered latent hierarchy that adapts seamlessly across compression levels and channel conditions. The full procedures are summarized in Algorithm~\ref{alg:inference} and Algorithm~\ref{alg:training}.
	
	\begin{algorithm}[t]
		\caption{Inference of the proposed prompt-conditioned PSI compression framework}
		\label{alg:inference}
		\begin{algorithmic}[1]
			\Require PSI tensor $\mathbf{T}$; side information $s=\{\mathrm{SNR}, \mathrm{ChanType}, r\}$
			\Ensure Reconstructed PSI $\hat{\mathbf{T}}$
			\State $\mathbf{X}_{\text{in}} \gets \mathrm{EmbedConv}(\mathbf{T})$ \Comment{Feature extraction}
			\State $(\mathbf{p}_f, \mathbf{p}_s) \gets g(s)$ \Comment{Prompt generation}
			\State $(\boldsymbol{\gamma}_\ell, \boldsymbol{\beta}_\ell, g_f) \gets \mathrm{Proj}_{\mathrm{FiLM}}(\mathbf{p}_f)$ \Comment{FiLM modulation}
			\State $\mathbf{S} \gets \mathrm{Proj}_s(\mathbf{p}_s)$; concatenate with $\mathbf{X}_{\text{in}}$ before attention \Comment{Soft prompts}
			\State $\mathbf{z} \gets f_{\mathrm{enc}}(\mathbf{T}; s)$ \Comment{Encoder forward pass}
			\State $\mathbf{z}_r \gets \alpha(r)\!\left(\mathbf{M}(r) \odot \mathbf{z}\right)$ \Comment{Variable-rate masking}
			\State $\tilde{\mathbf{z}} \gets \mathbf{H}_c \mathbf{z}_r + \mathbf{w}$ \Comment{Control-link transmission}
			\State $\hat{\mathbf{T}} \gets f_{\mathrm{dec}}^{\mathrm{DWCG}}(\tilde{\mathbf{z}})$ \Comment{DWCG decoding}
			\State \Return $\hat{\mathbf{T}}$
		\end{algorithmic}
	\end{algorithm}
	
	\begin{algorithm}[t]
		\caption{Training of the proposed prompt-conditioned PSI compression framework}
		\label{alg:training}
		\begin{algorithmic}[1]
			\Require Dataset $\mathcal{D}$; epochs $E$; batch size $B$; optimizer \textsc{Opt}; rate set $\mathcal{R}$
			\Ensure Trained parameters $\Theta$
			\For{$e = 1$ to $E$}
			\For{mini-batch $\{\mathbf{T}_b\}_{b=1}^{B} \subset \mathcal{D}$}
			\State Sample $(\mathrm{SNR}_b, \mathrm{ChanType}_b, r_b) \sim \mathcal{R}$
			\State Generate FiLM params $(\boldsymbol{\gamma}_{\ell,b}, \boldsymbol{\beta}_{\ell,b}, g_{f,b})$ and soft prompts $\mathbf{S}_b$
			\State $\mathbf{z}_b \gets f_{\mathrm{enc}}(\mathbf{T}_b; s_b)$
			\State Apply mask $\mathbf{M}(r_b)$; transmit $\tilde{\mathbf{z}}_b = \mathbf{H}_{c,b}\mathbf{z}_{r,b} + \mathbf{w}_b$
			\State $\hat{\mathbf{T}}_b \gets f_{\mathrm{dec}}^{\mathrm{DWCG}}(\tilde{\mathbf{z}}_b)$
			\EndFor
			\State $\mathcal{L} = \frac{1}{B}\!\sum_b \frac{\|\mathbf{T}_b - \hat{\mathbf{T}}_b\|_2^2}{\|\mathbf{T}_b\|_2^2}$
			\State $\Theta \gets \textsc{Opt}(\Theta, \nabla_\Theta \mathcal{L})$
			\EndFor
			\State \Return $\Theta$
		\end{algorithmic}
	\end{algorithm}

	\noindent\textbf{5) System Level Implications.}
	The proposed framework realizes a fully asymmetric yet harmonized autoencoder architecture. Prompt conditioning embeds instantaneous channel and rate awareness entirely at the encoder via the FiLM, soft prompt hierarchy, while the DWCG decoder performs efficient local reconstruction without requiring any side information. This division of labor, global adaptation at the BS and local refinement at the IRS controller, mirrors real hardware asymmetry, enabling scalable deployment across heterogeneous IRS platforms. Overall, the framework achieves low signaling overhead, high scalability, and real time adaptability, offering a unified, deployable, and energy efficient solution for variable rate PSI delivery in next generation IRS-aided wireless networks.

	\section{Simulation Results and Analysis}
	
	We conduct four sets of experiments to evaluate the proposed prompt conditioned PSI compression framework in alignment with its core design objectives.  
	Initially, we evaluate robustness by examining if prompt conditioning facilitates stable generalization across varying SNR levels and types of fading.  
	We assess rate flexibility by examining the capability of a single variable rate model to adaptively function across various compression ratios, potentially substituting multiple per-rate baselines.  
	We analyze scalability by examining the evolution of reconstruction performance as IRS size increases and verifying consistency with theoretical scaling trends.  
	We conduct an ablation study to evaluate the individual contributions of the encoder-side prompt conditioning and the DWCG-based decoder to overall accuracy and efficiency.  
	These experiments form a validation pipeline that demonstrates the proposed framework's effectiveness and computational efficiency within realistic deployment constraints.
	
	Results are presented as NMSE (dB) averages derived from 100 Monte Carlo trials, unless stated otherwise.  
	The model is optimized for 10,000 epochs using the Adam optimizer \cite{adam}, with an initial learning rate of $10^{-4}$ and a cosine annealing schedule.
	
	\subsection{Cross SNR and Cross Channel Adaptation}
	
	\begin{figure*}[t]
		\centering
		\subfloat[Rayleigh channel, $\mathrm{CR}=0.5$, $N=16$]{%
			\includegraphics[width=0.48\linewidth]{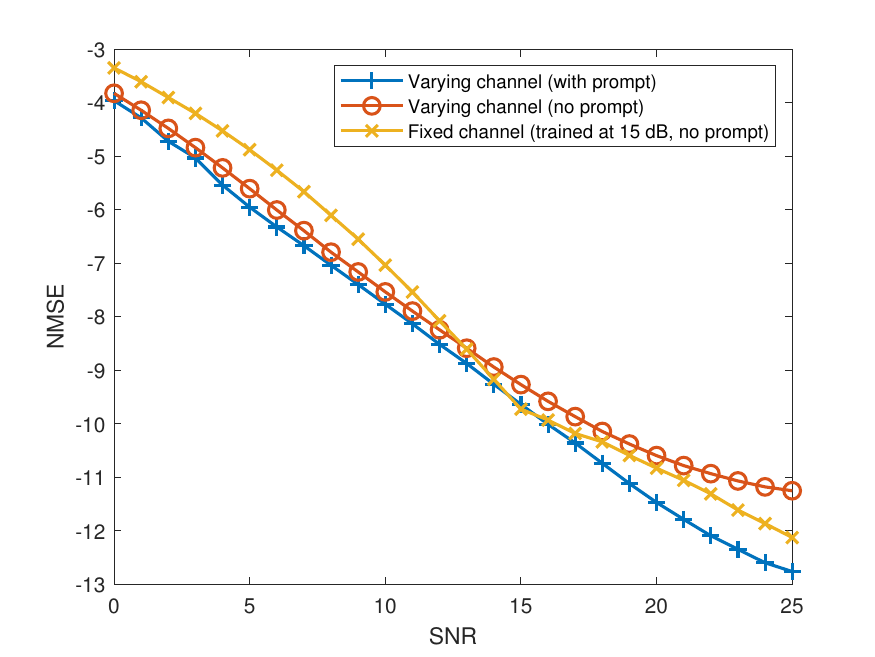}%
			\label{vsnrrayleigh}
		}
		\hfill
		\subfloat[Rician channel, $\mathrm{CR}=0.5$, $N=16$]{%
			\includegraphics[width=0.48\linewidth]{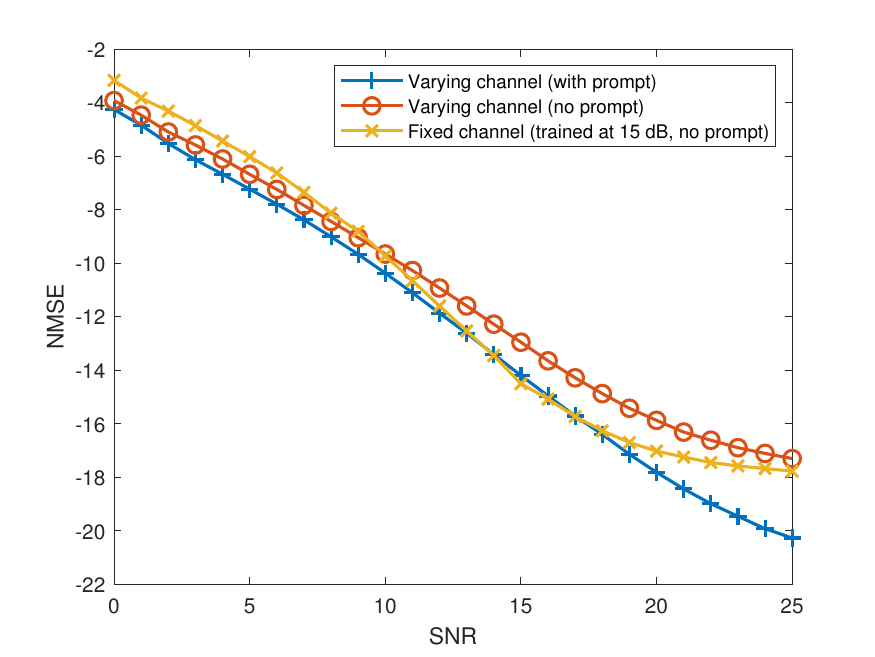}%
			\label{vsnrrician}
		}
		\caption{Cross SNR and cross channel adaptation performance of the proposed framework under Rayleigh and Rician fading.}
		\label{fig:E1_cross_snr}
	\end{figure*}
	
	This experiment evaluates the robustness of the proposed prompt conditioned encoder under varying channel statistics and SNR levels.  
	The IRS size is fixed at $N=16$ and the compression ratio at $\mathrm{CR}=0.5$, while the fading type alternates between Rayleigh and Rician environments.  
	Three variants are compared:  
	(i) \emph{Varying channel with prompt} (ours), trained jointly across SNR levels and fading types;  
	(ii) \emph{Varying channel without prompt}, trained under the same conditions but without prompt conditioning; and  
	(iii) \emph{Fixed channel without prompt}, trained solely on Rayleigh fading at $15$\,dB, serving as a matched yet nonadaptive reference.  
	
	Fig.~\ref{fig:E1_cross_snr} shows the resulting NMSE performance across SNR levels.  
	Under matched conditions, the fixed channel baseline attains slightly lower NMSE due to its environment specific optimization.  
	However, when tested under mismatched SNRs or unseen fading types, its performance deteriorates rapidly.  
	In contrast, the proposed prompt conditioned model maintains stable accuracy and consistently surpasses the non-prompt variant, with the performance margin widening as SNR increases.  
	This trend indicates that the encoder benefits more from adaptive prompt modulation when the input SNR is high and channel variability becomes more pronounced.  
	
	These results verify that the proposed hybrid prompt mechanism successfully integrates side information such as SNR and fading type into the encoder’s latent representation, thereby achieving strong cross domain generalization and stable reconstruction quality without retraining.
	
	\subsection{Rate Flexibility with Variable Compression Ratio Model}
	
	\begin{figure*}[t]
		\centering
		\subfloat[Rayleigh channel]{%
			\includegraphics[width=0.48\linewidth]{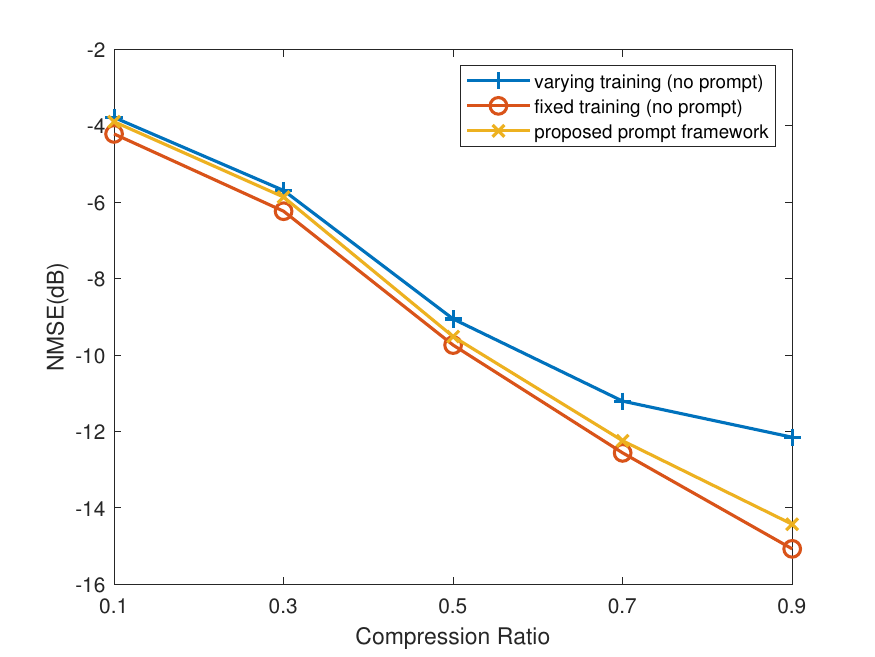}%
			\label{vcrn16ray15}
		}
		\hfill
		\subfloat[Rician channel]{%
			\includegraphics[width=0.48\linewidth]{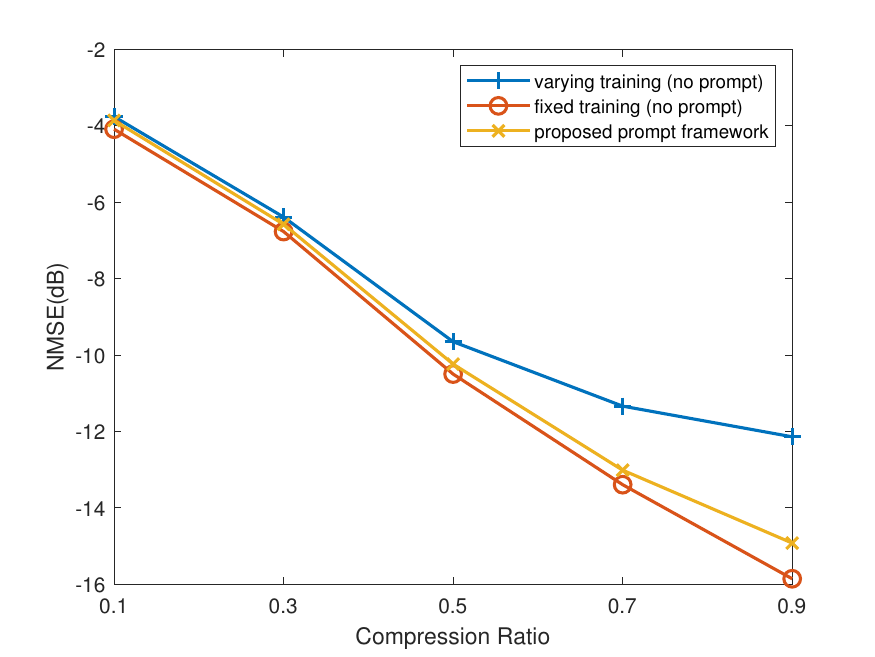}%
			\label{vcrn16rica10}
		}
		\caption{Performance under Rayleigh and Rician channels ($N{=}16$). 
			The proposed prompt conditioned framework achieves flexible rate adaptation across multiple compression ratios using a single unified model.}
		\label{fig:E2}
	\end{figure*}
	
	This experiment evaluates the framework’s ability to maintain high reconstruction quality across multiple compression ratios within a single variable rate model, thereby eliminating the need for separately trained per-rate networks.  
	In practical IRS control, the allowable signaling overhead may fluctuate with link quality, latency budget, or scheduling demands, making adaptive compression especially critical.  
	
	The IRS size $N = 16$, and both Rayleigh and Rician fading channels are considered. The proposed prompt conditioned model is trained once with randomly sampled compression ratios $r \in \{0.1, 0.3, 0.5, 0.7, 0.9\}$, while baseline models are independently optimized for each fixed rate.  
	The resulting NMSE performance is shown in Fig.~\ref{fig:E2}, where the $x$-axis represents the compression ratio and the $y$-axis depicts NMSE in dB.  
	Three configurations are compared:  
	(i) a non-prompt variable rate model trained across all rates,  
	(ii) multiple fixed-rate models trained individually, and  
	(iii) the proposed prompt conditioned variable rate model.  
	
	As observed in Fig.~\ref{fig:E2}, the proposed model achieves accuracy nearly identical to the best fixed rate networks, while consistently outperforming the non prompt variable rate baseline at all compression levels.  
	This demonstrates that prompt conditioning effectively regularizes the latent space, enabling the encoder to prioritize globally informative features under diverse compression budgets.  
	Moreover, the single model architecture achieves such flexibility without retraining or model switching, validating its practicality for adaptive, resource-constrained IRS deployments.  
	Overall, the prompt conditioned variable rate model delivers a unified and efficient solution that preserves performance optimality under dynamically varying compression requirements.
	
	\subsection{IRS Size Scalability and Theoretical Consistency}
	
	\begin{figure*}[t]
		\centering
		\subfloat[]{%
			\includegraphics[width=0.48\linewidth]{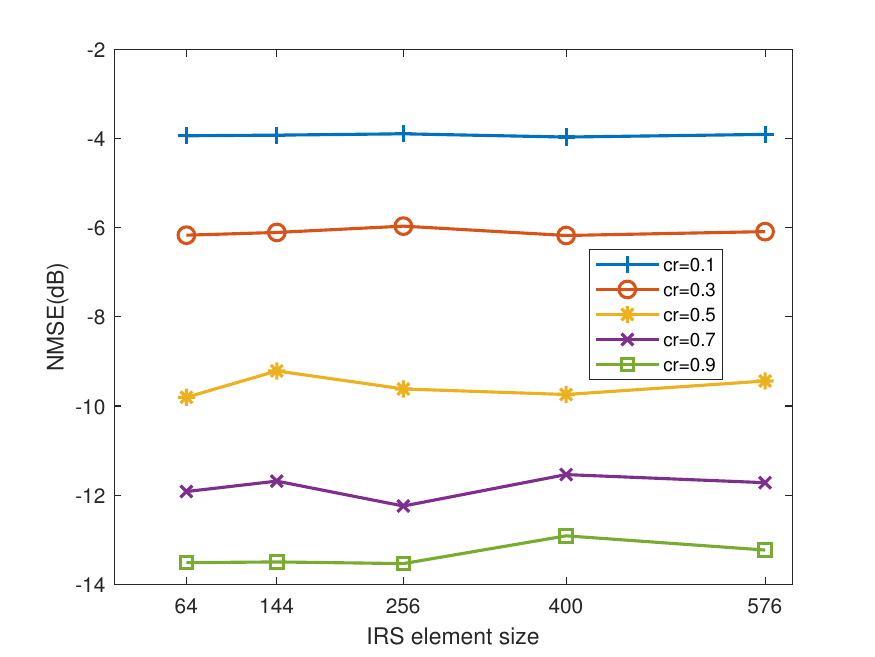}%
			\label{vnrayleigh}
		}
		\hfill
		\subfloat[]{%
			\includegraphics[width=0.48\linewidth]{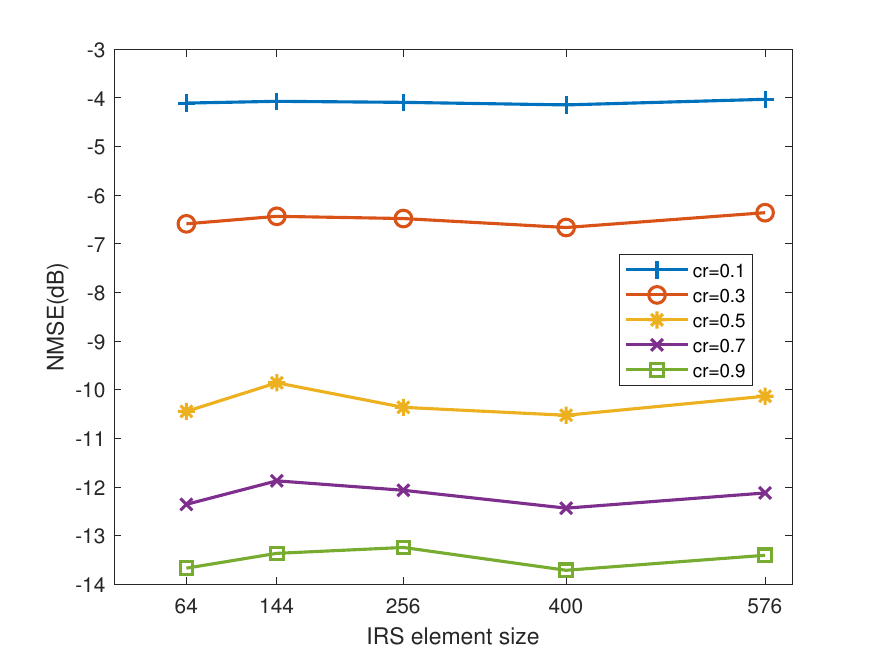}%
			\label{vnrician}
		}
		\caption{NMSE performance versus IRS element size under varying compression ratios ($\mathrm{CR}=0.1$–$0.9$) for (a) Rayleigh channel at $\mathrm{SNR}=15$\,dB and (b) Rician channel at $\mathrm{SNR}=10$\,dB, demonstrating stable scalability of the proposed model across IRS configurations.}
		\label{fig:E3}
	\end{figure*}
	
	This experiment investigates the scalability of the proposed prompt conditioned compression framework as the IRS size increases.  
	As future wireless systems evolve toward ultra-large reconfigurable surfaces, it is essential to verify that the compression mechanism generalizes across array sizes while maintaining reconstruction accuracy and computational efficiency.  
	To this end, the number of IRS elements is varied as \(N \in \{64, 144, 256, 400, 576\}\) under both Rayleigh and Rician fading, while keeping the model parameters and compression ratios fixed.  
	This configuration isolates the effect of array dimension, ensuring that any variation in NMSE directly reflects the framework’s scalability rather than differences in network capacity or training setup.
	
	As shown in Fig.~\ref{fig:E3}, the NMSE remains nearly constant as \(N\) increases, confirming that the learned feature compression is dimension agnostic and scales consistently with the array geometry.  
	Across all compression ratios, the performance trends align with theoretical scaling laws, indicating that the encoder’s latent representation effectively captures the spatial correlation structure of IRS elements.  
	Moreover, results under the Rician channel demonstrate that the framework preserves robustness in the presence of mixed line of sight and multipath fading components.
	
	Overall, the proposed prompt conditioned compression framework delivers consistent reconstruction accuracy and theoretical scalability across varying IRS sizes, validating its suitability for large scale IRS deployments in next generation wireless networks.
	
	\subsection{Ablation on Prompt Mechanism and Decoder Design}
	
	\begin{figure*}[t]
		\centering
		\subfloat[Rayleigh channel (20\,dB)]{%
			\includegraphics[width=0.48\linewidth]{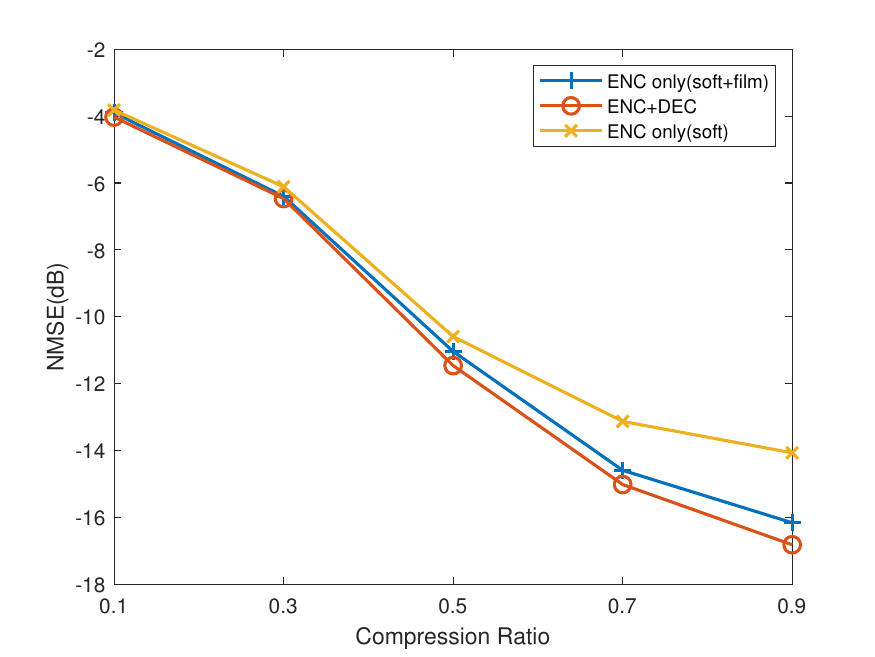}%
			\label{ablpromptray}
		}
		\hfill
		\subfloat[Rician channel (15\,dB)]{%
			\includegraphics[width=0.48\linewidth]{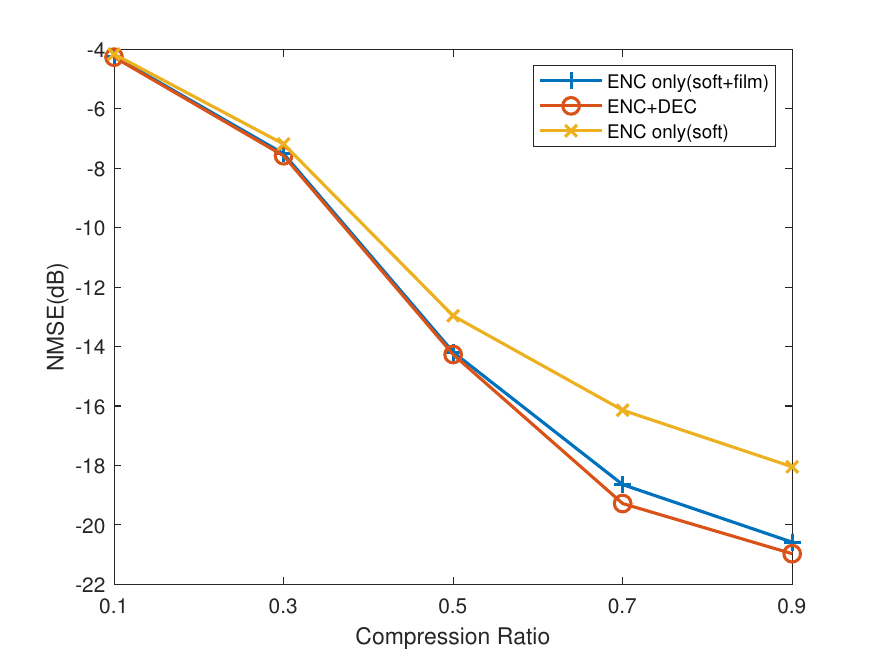}%
			\label{ablpromptrician}
		}
		\caption{Ablation study of the prompt conditioning mechanism under Rayleigh and Rician channels. 
			Comparison among (i) encoder only soft + FiLM prompts (proposed), (ii) encoder decoder joint prompting, and (iii) encoder only soft prompt. 
			Results indicate that encoder side film achieves the most favorable compression–reconstruction tradeoff.}
		\label{fig:E4_A1}
	\end{figure*}
	
	To further clarify the contribution of each component within the proposed compression framework, we conduct an ablation study focusing on the prompt-conditioning mechanism and decoder architecture.  
	Three variants are compared:  
	(i) encoder only prompting with both soft tokens and FiLM modulation (proposed),  
	(ii) encoder decoder joint prompting, and  
	(iii) encoder only soft prompts without FiLM modulation.  
	Experiments are performed under Rayleigh and Rician channels at representative SNR levels (20\,dB and 15\,dB, respectively) across varying compression ratios.  
	
	As shown in Fig.~\ref{fig:E4_A1}, the \emph{encoder only hybrid prompting} (soft + FiLM) achieves the most favorable compression reconstruction trade-off across both fading environments.  
	Introducing decoder side prompting (\textit{ENC+DEC}) yields marginal improvement only at extremely high compression ratios but at the cost of additional signaling and computational complexity.  
	In contrast, removing FiLM and retaining only soft prompting (\textit{ENC only (soft)}) leads to consistent NMSE degradation across all compression ratios, confirming that feature-wise statistical modulation is essential for stable encoding.  
	These results verify that concentrating prompt conditioning at the encoder effectively captures both semantic and statistical context while avoiding prompt-latent misalignment and over the air distortion, thereby enabling robust yet lightweight adaptation.
	
	\begin{figure}[t]
		\centering
		\includegraphics[width=1\linewidth]{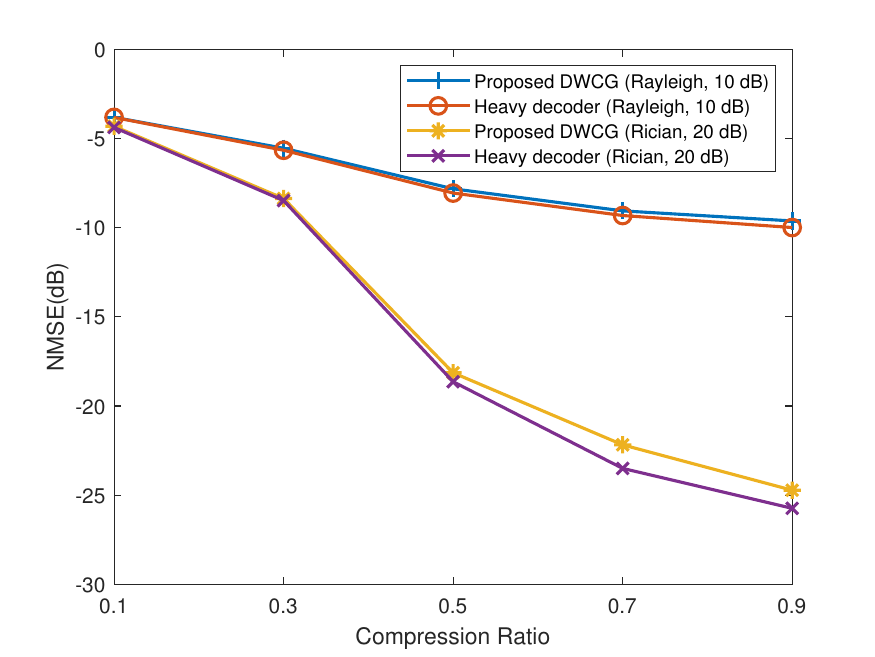}
		\caption{Decoder ablation: NMSE (dB) vs. compression ratio $r$ comparing the \textbf{DWCG decoder} and a \textbf{heavy attention-based decoder} under Rayleigh ($\mathrm{SNR}=10$\,dB) and Rician ($\mathrm{SNR}=20$\,dB). 
			Settings: $N{=}16$, variable rate prompt conditioned encoder, prompt-free decoder.}
		\label{fig:E4_A2}
	\end{figure}
	
	The impact of the lightweight DWCG decoder is further evaluated by replacing it with two heavier alternatives a standard convolutional decoder and a gated residual decoder adjusted to maintain comparable parameter counts.  
	As illustrated in Fig.~\ref{fig:E4_A2}, the DWCG structure achieves equal or superior NMSE performance while drastically reducing inference latency and computational complexity.  
	This demonstrates that DWCG provides an excellent tradeoff between reconstruction fidelity and efficiency, enabling real time operation on resource-constrained IRS controllers.
	
	\begin{table*}[t]
		\centering
		\caption{Complexity Comparison of Encoder Prompting and Decoder Designs}
		\label{tab:complexity_full}
		\begin{tabular}{lcccc}
			\toprule
			\textbf{Module} & \textbf{Variant} & \textbf{Params (×10\textsuperscript{6})} & \textbf{MACs (×10\textsuperscript{6})} & \textbf{Inference Time (ms)} \\
			\midrule
			\multirow{3}{*}{\textbf{Encoder}} 
			& Baseline (no prompt) & 10.91 & 197.33 & 4.07 \\
			& +Soft only & 11.12 & 207.41 & 4.38 \\
			& \textbf{+Soft + FiLM (proposed)} & \textbf{11.51} & \textbf{219.17} & \textbf{4.62} \\
			\midrule
			\multirow{2}{*}{\textbf{Decoder}} 
			& Baseline (attention-based) & 10.91 & 206.90 & 4.07 \\
			& \textbf{DWCG (proposed)} & \textbf{1.28} & \textbf{20.11} & \textbf{1.12} \\
			\bottomrule
		\end{tabular}
	\end{table*}
	
	As summarized in Table~\ref{tab:complexity_full}, incorporating FiLM and soft prompts at the encoder side increases both the parameter count and computation by less than $5\%$, confirming that hierarchical conditioning introduces negligible overhead relative to the baseline.  
	In contrast, replacing the attention-based decoder with the proposed DWCG architecture reduces the parameter count by more than $85\%$ and the multiply accumulate (MAC) operations by approximately $90\%$, leading to a $3.6\times$ reduction in inference time.  
	This encoder decoder asymmetry achieves an effective balance between adaptability and efficiency: the encoder remains flexible to channel variations through lightweight prompt modulation, while the decoder retains minimal complexity for real-time inference on embedded IRS controllers.  
	Overall, these results verify that the proposed hierarchical prompting and DWCG decoding together offer a scalable and deployable solution for large scale IRS systems, forming a practical foundation for future extensions toward distributed or multi-IRS deployments.
	
	\subsection{Summary and Discussion}
	
	The results across the four evaluation dimensions—cross SNR and cross channel adaptation, rate flexibility, scalability, and ablation—consistently validate the proposed prompt-conditioned PSI compression framework as effective and efficient.  
	The prompt-conditioned encoder exhibits robust generalization across diverse channels and SNR regimes, whereas the variable-rate design attains near-optimal performance within a cohesive network.  
	Scalability analyses demonstrate that the compression mechanism preserves theoretical consistency with increasing IRS size, while not introducing significant computational overhead.  
	The ablation results indicate that the encoder-side prompt embedding and lightweight DWCG decoder create a synergistic configuration that improves both adaptability and inference efficiency.  
	The findings collectively illustrate the feasibility of the proposed framework for large-scale, real-time IRS control, thereby establishing a basis for prompt-driven learning in wider communication and sensing applications.
	
	These experiments demonstrate how prompt-guided modulation connects static compression models with dynamic wireless environments.  
	The performance improvements result from the encoder's capacity to utilize contextual factors, including SNR, fading type, and compression ratio, to adaptively condition its latent representation, thereby achieving robustness and flexibility without the need for retraining.  
	However, several limitations persist: the current implementation depends on offline training and presumes known channel statistics for prompt generation.  
	The results establish a robust basis for scalable, prompt-aware control information processing in future intelligent surface networks.
	
	\section{Conclusion}
	
	This paper proposed a prompt-conditioned autoencoder framework for efficient PSI compression in IRS-aided wireless systems.  
	The integration of context-aware prompt embeddings into the encoder facilitates dynamic adaptation to diverse channel conditions, SNR levels, and compression ratios without necessitating retraining.  
	A lightweight DWCG decoder facilitates rapid and precise reconstruction with reduced complexity, rendering the overall design appropriate for real-time deployment at the network edge.
	
	Comprehensive experiments across four dimensions—robustness, rate flexibility, scalability, and ablation—demonstrated that prompt conditioning improves generalization in mismatched environments while offering precise rate control and significant scalability concerning IRS size.  
	The findings demonstrate that the proposed framework attains an optimal balance between accuracy and efficiency, successfully integrating model adaptivity with system practicality for next-generation reconfigurable wireless networks.
	
	Future research will advance this framework to encompass multi-IRS coordination and semantic-level control signaling, where prompt-based representations could integrate communication, control, and perception tasks within a unified adaptive learning paradigm.

\end{document}